# Monolithic integration of optically anisotropic GeSe-based films on GaAs by templated solid-phase epitaxy

Kira J. Martin[1], Autumn Y. Lee[1], Pranav Mahaadev[1], Pooja D. Reddy[1], Kelly Xiao[1], Tri Nguyen[1], Ashlee M. García[1], Aaron M. Lindenberg[1,2,3], Kunal Mukherjee[1*]

[1]Department of Materials Science and Engineering, Stanford University, Stanford, CA
[2]Stanford Institute for Materials and Energy Sciences, SLAC National Accelerator Laboratory, Menlo Park, California 94025
[3]Stanford PULSE Institute, SLAC National Accelerator Laboratory, Menlo Park, California 94025

*Corresponding Author: kunalm@stanford.edu

## Abstract

Layered IV–VI semiconductors such as GeSe exhibit strong in-plane optical anisotropy, making them promising candidates for polarization-sensitive photonic devices. However, realizing these properties in scalable platforms requires heteroepitaxial integration on technologically relevant substrates like GaAs. Direct growth of GeSe is complicated by its glass formation at low temperatures and high vapor pressure at elevated temperatures. To overcome this, we develop a method for ex-situ solid-phase epitaxy utilizing a SnSe buffer and offcut GaAs substrate to enable single-orientation crystalline GeSe films. Using polarized reflection measurements, we find that stabilizing a single-in-plane-orientation results in a 2× increase in anisotropic response between the armchair and zigzag directions. This work provides a new integration route to harness the anisotropic optical properties of GeSe and its alloys for polarization-sensitive technologies.

## I. Introduction

The integration of birefringent and dichroic semiconductors onto photonic and optoelectronic platforms can be leveraged for increased device functionality, such as polarization-sensitive detection, offering higher contrast imaging and polarization encoded communication.[1,2] While optical anisotropy can be introduced in isotropic materials by breaking symmetry through strain or nanofabrication, harnessing semiconductors with naturally anisotropic crystal structures is an elegant approach for monolithic integration.[3–7] Here, a single optically anisotropic semiconductor could replace discrete bulky components, simplify fabrication, and potentially yield even greater anisotropy in properties per unit length. These properties may be harnessed for devices such as polarization-sensitive photodetectors, polarizers, and optical switches.[8] Most demonstrations to date, however, rely on exfoliated or transferred crystals, limiting scalability and throughput compared with monolithically integrated thin films.[9–12] Single crystal heteroepitaxy of low-symmetry, non-cubic semiconductors on technologically relevant cubic substrates such as Si, Ge, GaAs, is a key challenge for integrating anisotropic electronic and optical properties on to technologically relevant device platforms. In general, the growth of low symmetry materials on high symmetry templates leads to rotational domains which average out the anisotropy in-plane.[7,13] With poor nucleation control, heteroepitaxy can even lead to polycrystalline films. Thus, studying how to epitaxially grow anisotropic semiconductors on isotropic substrates while preserving the full anisotropy of the film is a prerequisite for realizing these properties in scalable device platforms.

The prototypical example of an anisotropic semiconductor is black phosphorus (bP), a layered van der Waals (vdW) bonded material in the orthorhombic *Pnma* space group with a bandgap of 0.31 eV.[14] Its out-of-plane (OP) and in-plane (IP) structural differences yield different refractive indices, carrier mobility, and absorption coefficients along the three axes.[15] These properties result from a puckered structure with vdW interlayer bonding and mixed covalent-ionic bonding IP. Prior work with bP has shown large IP anisotropy in Raman scattering, transport properties, and optical absorption.[15,16] This anisotropy is a result of the low symmetry along the IP directions, often referred to as the armchair and zigzag directions.

While bP is hard to utilize for practical device applications due to its instability in air, germanium and tin selenides (GeSe, SnSe) are compelling alternatives since they share the same orthorhombic *Pnma* structure while offering substantially greater stability. GeSe (a=10.82 Å, b=3.83 Å, c=4.39 Å) and SnSe (a=11.50 Å, b=4.15 Å, c=4.43 Å) are semiconductors with bandgaps in the near infrared (1.1 and 0.9 eV respectively).[17–19] GeSe, in particular, is a glass former, suggesting potential for phase change functionality between an isotropic glass and anisotropic crystal.[20] GeSe and SnSe have largely been studied in their flake and polycrystalline form for thermoelectric applications.[21–23] These materials have also emerged as promising IP ferroelectrics.[24] Additionally, SnSe and GeSe have displayed large anisotropy in carrier mobility and optical responses.[25–28] Alloying GeSe and SnSe provides a route to tailor these anisotropic properties. Their IP anisotropic absorption has been harnessed for polarization-sensitive photodetectors with high polarization ratios and large spectral response ranges.[28–30] Moreover, the anisotropic carrier mobility allows for field effect transistors with direction dependent performance.[26] Much of the initial anisotropic property demonstration in these films have been through devices made on individual 'single crystal' flakes that are exfoliated and transferred or synthesized by physical vapor transport. The heteroepitaxy of these anisotropic semiconductors on mica or cubic MgO has been demonstrated with rotational domains.[31–33] SnSe has been grown fully anisotropic on a-plane sapphire.[33] Recently, we demonstrated that offcut GaAs (001) substrates can also suppress degenerate IP orientations in SnSe, yielding fully anisotropic epitaxial films.[7] Whether this approach generalizes to related *Pnma* semiconductors and their alloys with different lattice parameters or bonding character remains an open question.

In this work, we establish methods for heteroepitaxial growth of GeSe and SnGeSe alloys on GaAs substrates via molecular beam epitaxy (MBE). Unlike Sn-rich alloys, which can be grown epitaxially, the tendency of GeSe to form a glass rather than crystallize during deposition prohibits direct epitaxial growth, necessitating an ex-situ solid-phase epitaxy (SPE) approach in which an amorphous film is deposited and subsequently crystallized. With this approach, we achieve epitaxial thin films across the full SnSe-GeSe composition range. Structural characterization is used to study geometrical anisotropy and IP orientation, and polarization-sensitive reflection is used to characterize how geometrical anisotropy manifests itself as anisotropic optical properties in GeSe and SnGeSe thin films. Growth on on-axis substrates leads to the formation of two degenerate IP orientations, consistently limiting the IP anisotropy of our films. By using offcut substrates together with SPE, we stabilize a single-IP-orientation providing larger optical anisotropy.

## II. Results and Discussion

### a. Heteroepitaxy growth window of GeSe

GeSe films were deposited on GaAs (001) substrates in a MBE chamber with a compound GeSe effusion cell. Growth of GeSe is complicated by its high vapor pressure that limits sticking at higher growth temperatures and its glass forming tendencies that prevent direct crystalline deposition at lower growth temperatures.[34,20] Therefore, we first performed a growth temperature study to assess the feasibility of depositing crystalline GeSe with reasonable growth rates.

At a growth temperature of 250˚C, the reflection high energy electron diffraction (RHEED) pattern remained unchanged upon opening the GeSe shutter, retaining the (2×1) surface reconstruction of the GaAs surface even after

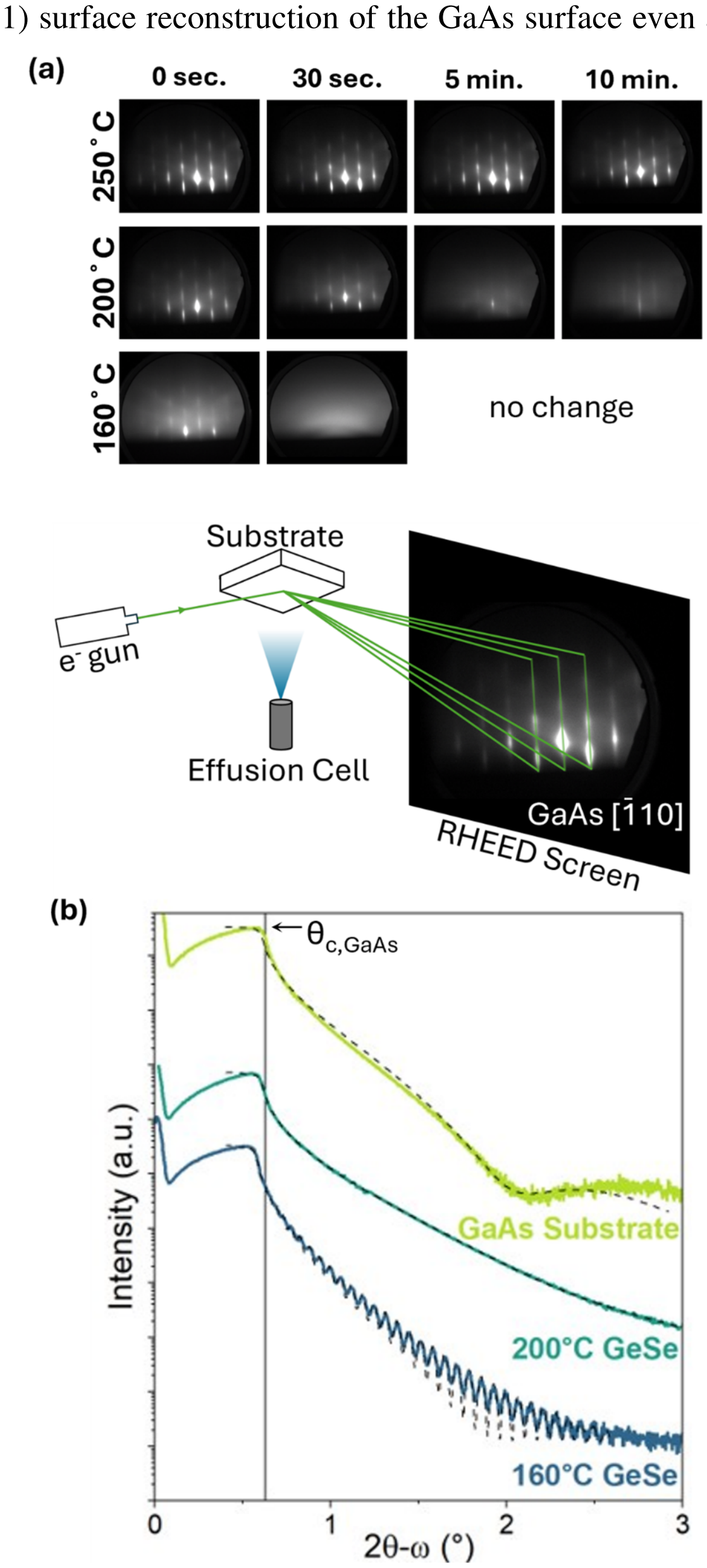

**Figure 1.** (a) RHEED pattern of [$\bar{1}$10] azimuth at the start, 30 seconds, 5 minutes, and 10 minutes into GeSe growth at three different growth temperatures. A schematic illustrates experimental setup. (b) XRR data and best fit of GaAs (001) substrate with native oxide and amorphous GeSe grown at a growth temperature of 200˚C and 160˚C. The curves are vertically offset for clarity. The vertical line marks the critical angle of the GaAs substrate.

10 minutes (Figure 1a). This suggests a near-zero sticking coefficient of GeSe on GaAs at 250˚ C. We next reduced the growth temperature to 200˚C to increase GeSe sticking. Upon opening the GeSe shutter, the RHEED pattern becomes partly hazy, suggestive of disordered or amorphous deposition. After 10 minutes the GaAs substrate pattern was still partially visible. RHEED is a surface sensitive technique that probes only the top several nanometers of material due to the grazing incidence angle[35], thus the sticking coefficient of GeSe appears still negligible at 200˚C. Finally, when the growth temperature was reduced to 160˚C, the substrate pattern disappeared as the RHEED became hazy after 30 seconds meaning the substrate surface was quickly buried under amorphous GeSe.

X-ray reflectivity (XRR) was used to characterize the growth rate of GeSe at each temperature as shown in Figure 1b. The 200˚C growth shows no clear thickness fringes but comparing it to an epi-ready GaAs substrate with the native oxide film, we can confirm GeSe has been deposited as the shift in critical angle suggests a decrease in the film's density. Fitting suggests the layer of GeSe has a thickness of ≤ 5 nm corresponding to a growth rate of ≤ 0.3 nm/min. In contrast, XRR of the 160˚C growth shows much clearer and closely spaced fringes due to the thicker GeSe film. The derived thickness of 120 nm corresponds to a growth rate of 2.4 nm/min which is nearly 8× higher than the 200˚C growth rate. Although we did not find a temperature window where GeSe crystalline growth and unity sticking regimes overlap, we chose a growth temperature of 160˚C for all the subsequent GeSe films due to the increased sticking and related growth reproducibility achieved.

**b. Solid-phase heteroepitaxy of GeSe on SnSe/GaAs**

Amorphous GeSe is inherently isotropic and must be crystallized to confer anisotropic properties. Atomic force microscopy (AFM) of the as-grown amorphous GeSe film reveals a very smooth (root mean square roughness of 0.9 nm) and continuous film in Figure 2d. We annealed the film *ex-situ* at 400˚C for 5 minutes and used a silicon wafer as a proximity cap to limit the desorption of Se and GeSe which are both volatile.[36] The capping method was not perfect as AFM of the annealed sample surface reveals droplets, likely of pure germanium (Ge) formed from Se desorption. A symmetric 2θ-ω X-ray diffraction (XRD) scan of this sample (referred to as "no buffer" in Figure 2c) only shows a small intensity GeSe (400) peak which is suggestive of a textured film with a preferred OP vdW axis orientation but low crystalline quality. Polarized optical microscopy (POM) in cross-polarized geometry is used to examine the spatial variation in optical properties of this textured film. As shown in Figure 2f, the GeSe film crystallized through spherulitic growth resulting in a polycrystalline film. The variation in color reflects the optical anisotropy of GeSe combined with a lack of control over OP and IP orientation during crystallization. Overall, the GaAs surface is unable to appreciably template the crystallization of GeSe.[37]

We find that a SnSe buffer better templates the crystallization of GeSe. SnSe shares the same orthorhombic *Pnma* structure as GeSe, but is not prone to glass formation during growth, allowing direct epitaxy. And in previous work, we showed similar growth conditions result in double-variant epitaxial SnSe films on exact GaAs (001), the out-of-plane epitaxial relationship is SnSe[100]||GaAs[001] for both variants, and the IP relationships are SnSe[010]||GaAs[110] and SnSe[001]||GaAs[110].[38] Therefore, we fabricated a sample, the structure diagramed in Figure 2b, with a 50 nm buffer layer of SnSe first grown on the GaAs substrate at 300˚C to help template the *ex-situ* crystallization of GeSe. A 100 nm layer of amorphous GeSe was deposited on top of the buffer layer. The root mean square (RMS) surface roughness of the amorphous film has increased to 1.7 nm as it mirrors the roughness of the SnSe buffer layer below. SnSe tends to grow via spiral growth mode from island coalescence which creates the lumpy background texture seen in the AFM in Figure 2g.[39] After annealing, the XRD scan (referred to as "SnSe buffer" in Figure 2c) shows high intensity peaks for the {100} family of planes for both the as-grown crystalline SnSe and solid-phase crystallized GeSe. Additionally, AFM of the sample post-annealing shows an increased RMS surface roughness of 2.8 nm due to Ge crystallites from Se desorption and a higher density of protruding islands. In addition to OP orientation, control of the IP crystallographic orientation of the crystallized GeSe is critical to access the complete anisotropy. Figure S1 shows a reciprocal space map (RSM) around the (224) reflection of GaAs. We index the (820) and (802) peaks of the double-variant SnSe film, that is—two orthogonal IP orientations rotated by 90˚—in line with our previous work.[38] The (820) peak is associated with the zigzag direction while the (802) peak is associated with the armchair direction. Remarkably, the GeSe film shows the same tendency, validating our hypothesis that a SnSe buffer can template both the OP and IP orientation of the solid-phase crystallized GeSe. In the film there are grains in which the armchair direction is aligned along the [110] direction of GaAs and others with the zigzag direction along the [110]. These grains can be visualized in Figure 2i. POM reveals two distinct family of grains distinguished by difference in hue and intensity which arises from the IP birefringence and dichroism of GeSe. Notably, the unannealed sample exhibits a similar microstructure in POM (Figure S2), where the optical contrast is due to the underlying crystalline SnSe buffer. The similarity in microstructure suggests a high density of GeSe nuclei form during crystallization and inherit the IP orientation established by the SnSe buffer.

The presence of two orthogonal IP orientations is not ideal as it reduces the effective IP anisotropy of the film as the IP properties become a weighted average of the properties along both the zigzag and armchair directions. We quantify this double-variant population using XRD. When the sample is azimuthally rotated by 90˚ (i.e. in φ), there is a change in relative peak intensities between the (820) and (802) peaks which, after accounting for structure factor differences, suggest an uneven volume fraction of the two IP orientations. Integrating under the curve of the film peaks and correcting for differences in structure factors, we find an 80:20 volume fraction ratio of the two IP orientations for both the SnSe and GeSe layers. Thus, the SnSe is exactly templating the GeSe, and achieving single-variant GeSe requires a single-variant SnSe buffer.

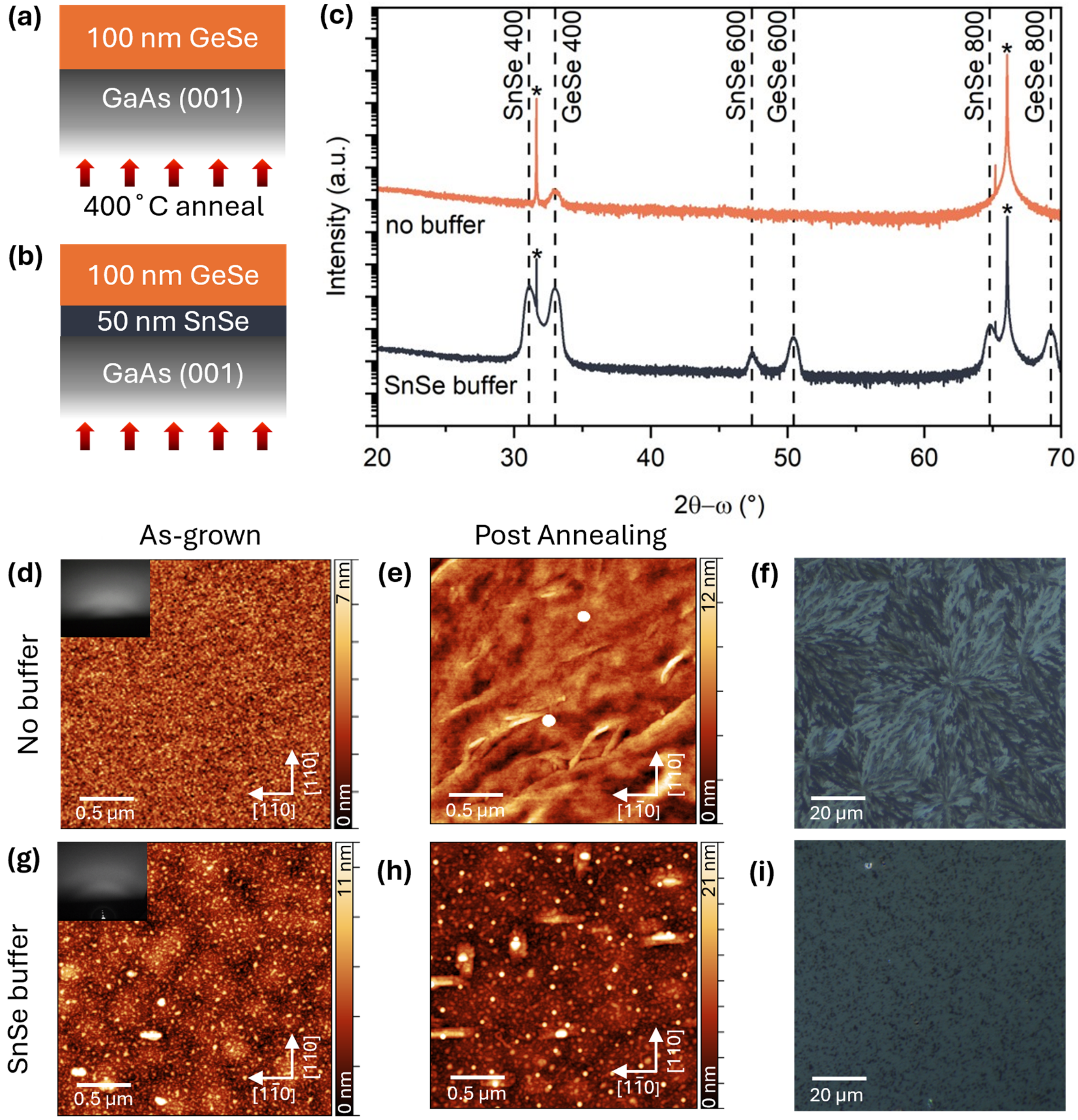

**Figure 2.** (a) Schematic of annealed GeSe on GaAs (001). (b) Schematic of annealed GeSe with crystalline SnSe buffer on GaAs (001). (c) Open detector symmetric 2θ-ω scans of both films post annealing. The scans are vertically offset for clarity. The asterisk is used to denote GaAs substrate peaks. (d) AFM scan of as-grown amorphous GeSe with inset of RHEED pattern from the end of the growth and (e) post annealing. (f) Polarized optical microscopy of annealed GeSe film to highlight microstructure and optical anisotropy. (g) AFM scan of as-grown amorphous GeSe with crystalline SnSe buffer with inset of RHEED pattern from the end of the growth and (h) post annealing. (i) Polarized optical microscopy of annealed GeSe film on SnSe buffer reveals different hues and microstructure.

We have previously demonstrated that SnSe on a GaAs (001) substrate with a 4˚offcut in the <111>B direction results in a single-IP-orientation. This is hypothesized to occur due to the zigzag edge preferentially aligning to the GaAs step edge.[7] We find that this method can be extended to solid-phase epitaxy by growing an identical layer stack of 100 nm amorphous GeSe deposited on a 50 nm crystalline SnSe buffer on a 4˚offcut (001) GaAs substrate. The surface morphology is altered when we switch to an offcut substrate; as seen in Figure 3b, a wave-like texture is formed as the SnSe nucleates and grows out from the step edges. This morphology is maintained even after GeSe crystallization, but we note the presence of pinholes likely formed from Se desorption. After annealing, the RSMs in Figure 3c show only the (820) GeSe peaks at φ=0˚and only the orthogonal (802) peaks at φ=90˚ confirming a single-IP-orientation, mirroring the structure of the SnSe buffer. Note that the GeSe and SnSe (802) peak positions in the φ=90˚ scan are rotated in the $Q_x$-$Q_z$ plane as the SnSe buffer is tilted ~1.5˚ with respect to the substrate due to details of nucleation on an offcut substrate.[7] Overall, our approach of a buffer + SPE successfully stabilizes single-orientation films of GeSe on GaAs, an otherwise difficult film to heteroepitaxially integrate directly as it is a glass former at low temperature and has a high vapor pressure at high temperature.

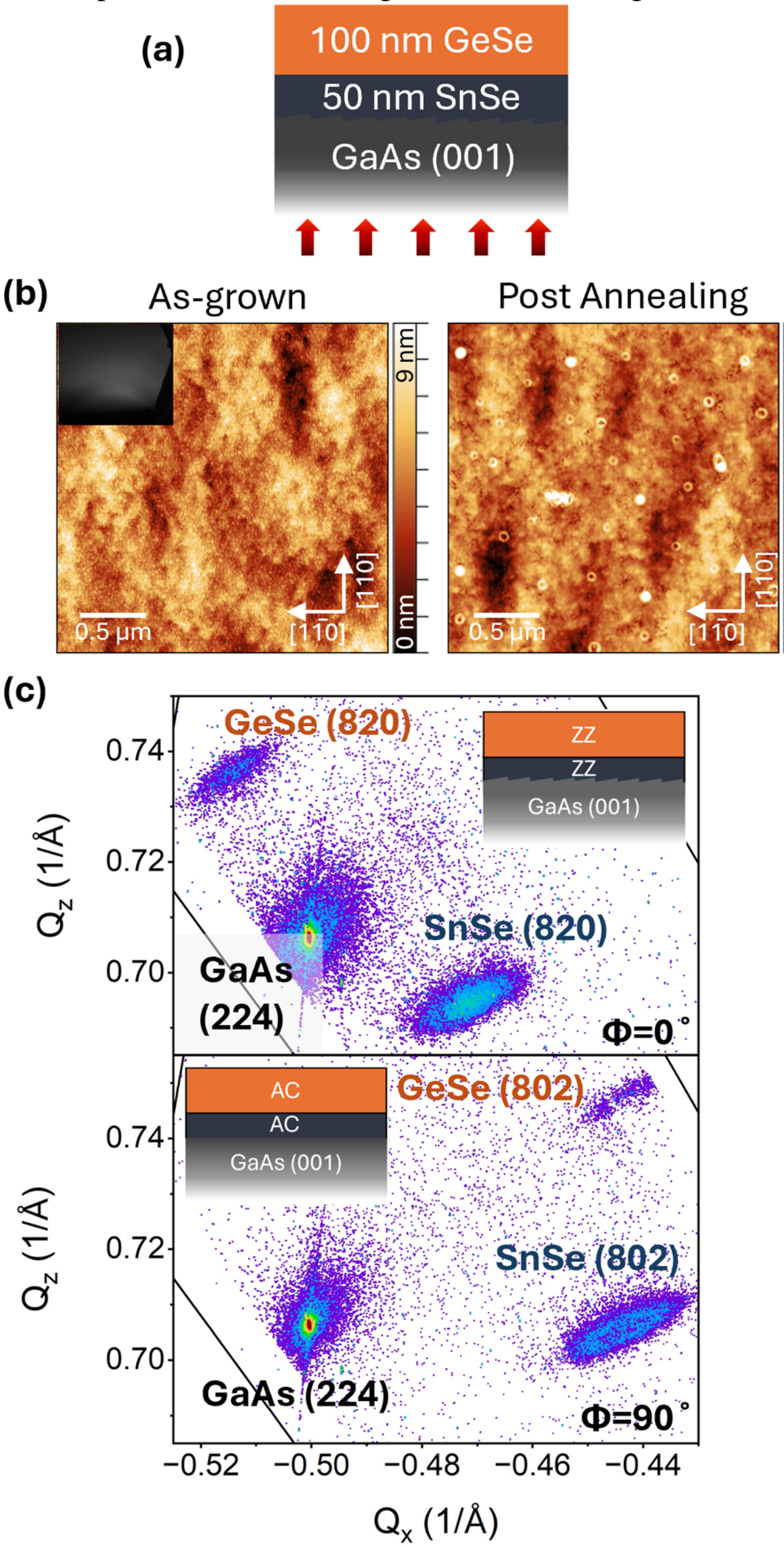

**Figure 3.** (a) Schematic of amorphous GeSe with crystalline SnSe buffer on 4˚offcut GaAs (001). (b) AFM scan of as-grown amorphous GeSe and post annealing. (c) RSMs of GeSe film with SnSe buffer grown on offcut GaAs (001) post annealing at azimuths φ=0˚ and 90˚. Schematics highlight how the GeSe layer maintains the IP orientation of the SnSe buffer after crystallization.

To assess how IP orientation control of GeSe impacts optical anisotropy, we performed polarization-sensitive reflection measurements with an 800 nm laser. We note that these measurements probe a multilayer structure of GeSe, SnSe, and GaAs. In Figure 4, both GeSe samples exhibit modulation, with much larger modulation observed for the single-IP-orientation film grown on offcut GaAs. The data was fit to a sinusoidal function with a period of 180˚ where the maximum and minimum reflected power correspond to reflection along the zigzag and armchair direction respectively. The ratio between these extremes ($\frac{I_{max}-I_{min}}{I_{min}}$), was used as a metric for IP optical anisotropy. The double-IP-orientated GeSe film exhibits a ratio of 0.04, while the single-IP-orientated sample has a ratio of 0.08. The single-IP-orientated GeSe has a 2× enhancement in optical anisotropy compared to the double-IP-orientated GeSe. This is in good agreement with expectations that eliminating the 20% volume fraction of orthogonal grains would cause a 1.7× increase in the amplitude. Additionally, with a 1030 nm laser we find the double-IP-orientated GeSe displays a ratio of 0.04, while the single-IP sample exhibits a ratio of 0.06 (Figure S3). The decreases in anisotropy ratio at 1030 nm suggests a decrease in Δn, Δk, or both; however, the wavelength dependence of these quantities remains uncertain in the literature.[9,40–42]

### c. Extension to SnGeSe alloys

With successful integration of GeSe, we now combine these growth methods to synthesize epitaxial films of SnGeSe alloys as a method to tune the geometrical anisotropy and spectral response through changes in composition. Bandgap engineering through composition is particularly important to consider as the material's anisotropic response is highly wavelength dependent.[9,28] The alloy space was swept by keeping the SnSe beam equivalent pressure (BEP) constant and varying the GeSe BEP to increase the Ge composition. The growth temperature of the SnGeSe alloy layer was kept as 160˚C to help with GeSe incorporation. While bulk GeSe is fully miscible in SnSe, the low growth temperature is expected to create a crystalline to amorphous transition once a critical germanium content is surpassed.[17] Crystalline alloys were grown up to a composition of $x_{Ge}$=0.39. At $x_{Ge}$=0.45 and above, the large GeSe flux resulted in amorphous alloys which were later crystallized *ex-situ* following the thermal annealing method discussed previously. This identifies a critical germanium content (between $x_{Ge}$=0.39–0.45), beyond which amorphous alloys are preferentially formed due to kinetic limitations at this low growth temperature.

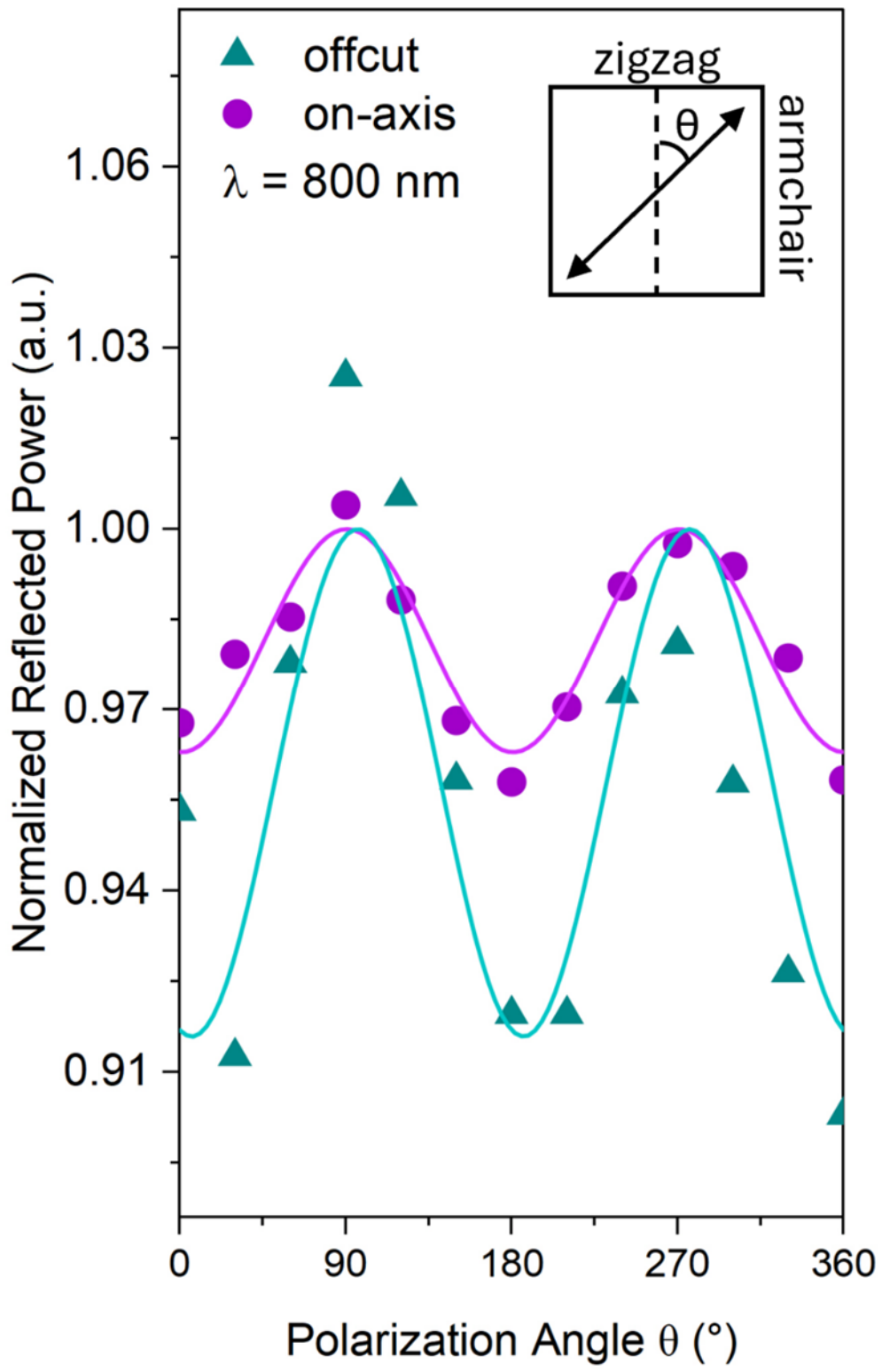

**Figure 4.** Normalized reflected power vs polarization angle for GeSe with single IP orientation (offcut) and double IP orientation (on-axis).

A 300˚C SnSe buffer layer is still necessary during crystalline alloy growth for out-of-plane orientation control at lower temperatures on exact GaAs (001). SnGeSe alloys grown directly on GaAs at 160˚C result in mixed orientations such as the (311) being stabilized (Figure S4). These off-(100) orientations are likely associated with direct covalent bonds between the film and GaAs due either to poor selenium passivation of the substrate or limited kinetics of the adatoms. Mixed-orientations were also observed when a thinner 10 nm SnSe buffer layer was used. The buffer layer was likely not yet fully coalesced, allowing some SnGeSe to directly bond with the GaAs substrate.

The IP and OP lattice constants were calculated using the SnGeSe peak positions from the GaAs (224) reciprocal space maps shown in Figure 5a. As the Ge content increases, the SnGeSe peaks shift away from the SnSe buffer peaks due to the change in lattice constants. The alloy compositions were estimated based on the shifting of the (800) SnGeSe OP peak position in symmetric 2θ-ω XRD scans and bulk alloy lattice constant trends from Krebs (Figure S5a).[17] Note the presence of the orthorhombic (820) and (802) peaks for both the SnSe buffer and SnGeSe alloy. This signifies the presence of two IP orientations in all SnGeSe alloys regardless of composition. Figure 5b summarizes how the lattice constants change across the alloy space. The OP lattice constant *a* decreases with increased Ge content following the expected bulk trend. As the Ge content increases the IP lattice constants *b* (zigzag) and *c* (armchair) diverge from each other, representing an increase in structural IP anisotropy. The experimental thin film lattice constant trends are slightly offset from the bulk trends due to strain from differences in thermal expansion coefficients between the SnGeSe film and GaAs substrate. As the sample is cooled down from the growth temperature to room temperature, the film expands or contracts depending on the sign of its thermal expansion coefficients, but the IP lattice constants are constrained by the GaAs substrate due to thermal expansion mismatch. The zigzag lattice constants are slightly tensile strained due to their positive thermal expansion coefficient. On the other hand, the negative thermal expansion coefficient of the armchair direction makes it compressively strained.[43] This causes tensile strain to also manifest in the out-of-plane *a* lattice constant roughly following Poisson's ratio. The structural evolution across alloy space is further supported by Raman spectroscopy, where the characteristic Raman peaks shift towards higher wavenumbers as Ge content increases (Figure S5b). The observed variation in structural anisotropy across the alloy space makes SnGeSe alloys intriguing for tuning anisotropic properties. However, it is important to note that SnSe and GeSe exhibit opposite transmission and absorption trends along the zigzag and armchair directions.[25] Consequently, alloying may enhance or suppress optical anisotropy and may even result in an isotropic optical response at a specific composition.

Furthermore, we determined the growth temperature of 160 ˚C is near unity sticking as we see a linear trend between Ge content and the ratio of GeSe BEP to total BEP during growth regardless of growth rate (Figure S5c). This is ideal for alloy growth as it makes the composition easier to control. Rocking curves were used to probe how the structural quality of the alloy layer changes with composition and growth rate in Figure S5d. We find that an increased growth rate has a stronger effect on the full width at half-maximum (FWHM) than alloying. Additionally, the FWHM of pure GeSe is on par with the as-grown crystalline SnSe sample, but the $x_{Ge}$=0.45 SnGeSe FWHM is much larger suggesting an increased barrier to epitaxial crystallization in the alloys.

We then explored the effect of miscut substrates on as-grown crystalline alloys by regrowing the $x_{Ge}$=0.18 alloy with a SnSe buffer on 4˚ offcut (001) GaAs. The SnSe buffer remained single-IP-orientated, but surprisingly the SnGeSe layer is double-IP-orientated (Figure S6a). The formation of a second IP orientation in the alloy is believed to be due to the lower growth temperature as we also observe the stabilization of two IP orientations in pure SnSe grown at 160°C on offcut GaAs despite the presence of a 300°C SnSe buffer (Figure S6b). We quantify the change in optical anisotropy with composition on the $x_{Ge}$=0.18 samples using the polarized reflection method detailed previously. At 1030 nm, the on-axis sample exhibits an anisotropy ratio of 0.06, while the offcut sample has a ratio of 0.07. There is only a minimal increase in anisotropy (1.2×) from using the offcut substrate because the alloy still has two IP orientations. At 800 nm, no distinct oscillations outside of the measurement's noise level are observed for both the SnGeSe samples grown on exact and offcut GaAs. This is directly opposed to the GeSe results which had a stronger anisotropic response at 800 nm. We attribute this difference to bandgap shifting, caused by alloying, changing the birefringent (Δn) and dichroic response (Δk) of the material. Overall, alloying offers a potential route for modulating optical anisotropy to tune device response across wavelength, but low-temperature alloy growth does not yet preserve single-variant orientation, limiting the attainable optical anisotropy.

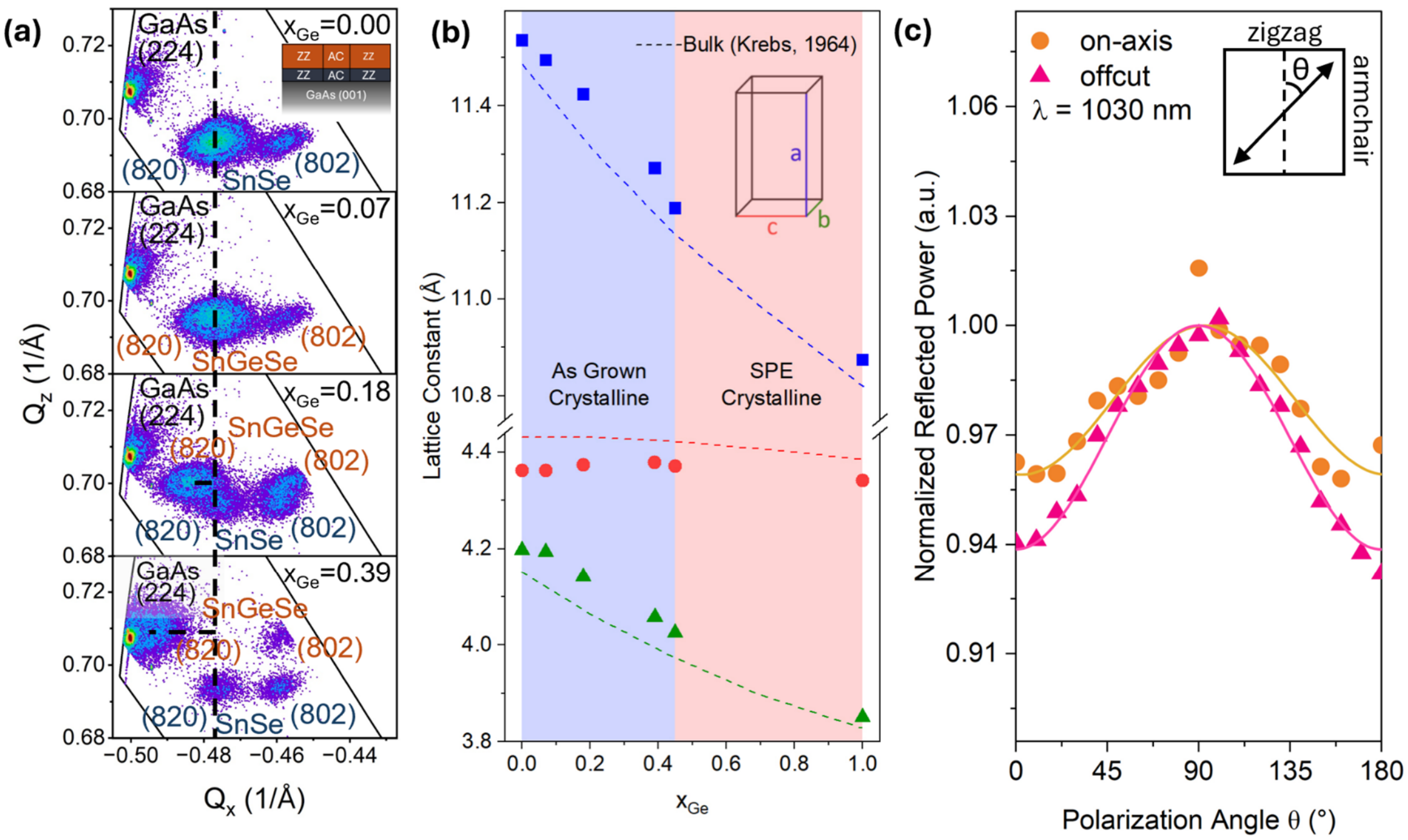

**Figure 5.** (a) RSMs of as grown crystalline SnGeSe alloys with SnSe buffer on GaAs (001). Two peaks for the SnSe buffer and SnGeSe film corresponding to two in-plane orientations. (b) Lattice constant as a function of Ge composition compared to bulk alloy trends. Lattice constants were determined from (820) and (802) alloy peaks in RSMs. (c) Normalized reflected power vs polarization angle for $x_{Ge}$=0.18 SnGeSe films grown on on-axis and offcut GaAs.

## III. Summary and Conclusions

In summary, we have developed a synthesis route for GeSe and SnGeSe films on (001) GaAs using MBE and solid-phase epitaxy. Due to the high vapor pressure and glass forming behavior of GeSe, we found a high-temperature SnSe buffer is necessary to template subsequent low-temperature deposition and SPE of GeSe and SnGeSe alloys. Additionally, for GeSe we demonstrated the ability of offcut substrates to stabilize layered structures with a single-IP-orientation via SPE. This development allows us to access the full IP optical anisotropy of GeSe. Further investigation is required to stabilize a single-IP-orientation during low temperature SnGeSe growth and characterize the composition-dependent anisotropic optical response across a broad spectral range. We have shown that epitaxial orientation control of anisotropic vdW semiconductors on GaAs is achievable through a combination of SnSe buffer layer templating, solid-phase epitaxy, and offcut substrate engineering. This has potential implications for the monolithic integration of polarization-sensitive devices and on-chip polarimetry components, with alloy composition providing an additional handle to tune the spectral range of the anisotropic response.

## IV. Methods

### a. Thin film Synthesis

Films were grown in a Riber Compact 21 chalcogenide MBE using compound SnSe, GeSe, and PbSe sources. For some growths the elemental valved Se cracker cell was used to supply a Se overpressure. The initial GeSe growth temperature series was grown on semi-insulating epi-ready GaAs (001) substrates. The native oxide layer was desorbed in the MBE under a Se overpressure at a pyrometer temperature of 570˚C for 10 minutes. RHEED was used to confirm the oxide was desorbed as the surface reconstruction became a strong (2×1) pattern typically seen for a Se-terminated GaAs (001) surface.[44,45] GeSe was then deposited at various temperatures using a BEP of $3.0\times10^{-7}$ Torr.

Once a growth temperature of 160˚C was chosen, samples were grown on semi-insulating GaAs (001) substrates with a pristine homoepitaxial layer. These substrates were prepared in a Veeco Gen III MBE. The oxide of semi-insulating epi-ready GaAs (001) substrates was thermally desorbed before 100 nm of homoepitaxial GaAs was grown under an As overpressure. An amorphous As capping layer was then deposited to preserve the growth surface from oxidation. Out-of-vacuum, these substrates were cleaved into 1 $cm^2$ pieces which were indium bonded onto molybdenum platens and transferred into the Riber MBE.

The As cap was thermally desorbed at 420˚C for 10 minutes revealing the (2×1) surface reconstruction of the GaAs substrate below. At the same temperature, the surface was treated using a PbSe BEP of $3.0\times10^{-7}$ Torr. PbSe does not stick at this high temperature, but this step has been shown to promote the epitaxial growth of IV-VI semiconductors.[46] The SnSe buffer layer was then grown at 300˚C with a BEP of $3.0\times10^{-7}$ Torr. The SnGeSe alloys were then grown at 160˚C by varying the ratio of SnSe and GeSe BEP's.

*Ex-situ* annealing of amorphous films was done on a hot plate at 400˚C for 5 minutes and the sample was capped with a Si wafer to minimize Se desorption from the surface.

### b. Structural Characterization

The surface morphology was studied using the Park NX-10 AFM with a NCS15 tip in tapping mode. X-ray diffraction was performed on a PANalytical Empyrean instrument with Cu-Kα radiation. X-ray reflectivity was measured with mirror incident optics and parallel plate collimator as the receiving optics. Symmetric 2θ-ω scans and rocking curves were measured in an open detector configuration. Triple axis configuration symmetric 2θ-ω scans were taken around the GaAs 004 peak to see the splitting of the SnSe and SnGeSe 800 peaks. The position of the SnGeSe 800 peak was used to calculate alloy compositions. For these calculations it was assumed all alloys were strained the same amount as pure SnSe. Reciprocal space maps in grazing incidence geometry were used to determine lattice constants and in-plane film orientation.

### c. Optical Characterization

The HORIBA Scientific LabRAM HR Evolution spectrometer was used to measure Raman spectra with a 633 nm laser with a 600 gr/mm grating and 300 second accumulation.

Cross-polarized optical microscopy images were taken on a Nikon LV100ND microscope in reflection mode with white light.

Polarization-dependent reflectivity of GeSe samples were measured in a microscope set up with an Olympus LCPlan N 20×/0.45 IR objective. The 50 femtosecond, 800 nm, light from a Ti:Sapphire source (Femtolaser XL500) at 5.1 MHz passed through a controllable half-wave plate for polarization control and was focused to a ~5 µm spot by the objective at normal incidence. The beam was attenuated using an ND filter wheel, resulting in a fluence of approximately 2 $mJ/cm^2$. The reflected light was collected by the same objective and redirected by a beamsplitter to a free-space biased silicon photodetector (Thorlabs DET10A). Each measurement was normalized by a reference measured in parallel to account for power fluctuations in time.

GeSe and SnGeSe samples were measured in a microscope setup using linearly polarized light of wavelength 1030 nm. Light from a fiber-laser based chirped pulse amplifier (Calmar Cazadero FLCPA-02USLC, 1.28 MHz rep. rate) was passed through a 50:50 beamsplitter (Thorlabs BSW27). Variations in incident power were monitored using a free-space biased silicon photodiode (Thorlabs PDA100A) on the reflected arm, and later normalized. The transmitted arm was passed through a motorized half-wave plate for polarization control. It was then focused to a ~43 um spot on the sample using an objective (Newport F-L20 8.6 mm/0.5). Power incident on the sample was adjusted using the current-seed of the laser, resulting in a fluence of 56 uJ/cm^2. The reflected light, collimated by the same objective, was redirected by the beamsplitter to another free-space biased Si photodiode (Thorlabs PDA100A2). To eliminate contributions from stray back-reflections of the waveplate and objective, a background-subtraction routine was adopted. Finally, the reflectivity of the sample was divided by the reflectivity of isotropic Si, to account for the anisotropy of any optics upstream.

**Supporting Information**
The data supporting this article have been included as part of the Supplementary Information.

**Acknowledgments** We gratefully acknowledge support via the NSF CAREER award under Grant No. DMR-2036520 and the UC Santa Barbara NSF Quantum Foundry funded via the Q-AMASE-I program under Award No. DMR-1906325 for MBE synthesis and structural characterization. Part of this work was performed at nano@stanford RRID:SCR_026695. K.J.M. gratefully acknowledges this material is based upon work supported by the National Science Foundation Graduate Research Fellowship Program under Grant No. DGE-2146755. Any opinions, findings, and conclusions or recommendations expressed in this material are those of the author(s) and do not necessarily reflect the views of the National Science Foundation. A.Y.L., P.M., and A.M.L. acknowledge support from the US Department of Energy, Office of Basic Energy Sciences, Division of Materials Sciences and Engineering (Contract No. DE-AC02-76SF00515)

**Author Declarations**

Conflict of Interest

The authors have no conflicts to disclose.

Author Contributions

Kira J. Martin: Formal analysis (equal); Investigation (lead); Methodology (lead); Validation (lead); Writing – original draft (equal). Autumn Y. Lee: Investigation (supporting); Methodology (supporting); Writing – review & editing (equal). Pranav Mahaadev: Investigation (supporting); Methodology (supporting); Writing – review & editing (equal). Pooja D. Reddy: Investigation (supporting); Writing – review & editing (equal). Kelly Xiao: Investigation (supporting); Writing – review & editing (equal). Tri Nguyen: Investigation (supporting); Writing – review & editing (equal). Ashlee M. García: Investigation (supporting); Writing – review & editing (equal). Aaron M. Lindenberg: Methodology (supporting); Supervision (supporting); Writing – review & editing (equal). Kunal Mukherjee: Conceptualization (equal); Formal analysis (supporting); Funding acquisition (lead); Methodology (supporting); Supervision (lead); Validation (supporting); Writing – original draft (equal).

**Data Availability**

The data that support the findings of this study are available from the corresponding author upon reasonable request.

**Supplementary Material**

**Monolithic integration of optically anisotropic GeSe-based films on GaAs by templated solid-phase epitaxy**

Kira J. Martin[1], Autumn Y. Lee[1], Pranav Mahaadev[1], Pooja D. Reddy[1], Kelly Xiao[1], Tri Nguyen[1], Ashlee M. García[1], Aaron M. Lindenberg[1,2,3], Kunal Mukherjee[1*]

[1]Department of Materials Science and Engineering, Stanford University, Stanford, CA

[2]Stanford Institute for Materials and Energy Sciences, SLAC National Accelerator Laboratory, Menlo Park, California 94025

[3]Stanford PULSE Institute, SLAC National Accelerator Laboratory, Menlo Park, California 94025

*Corresponding Author: kunalm@stanford.edu

**S1: In-plane orientation of annealed GeSe with SnSe buffer**

Asymmetrical reciprocal space maps (RSM) around the GaAs (224) reflection were used to probe the in-plane (IP) orientation of the SnSe and GeSe layers. The presence of the (820) and (802) reflections for both the SnSe buffer and crystallized GeSe indicates double-IP-orientations in both layers. The (802) reflection corresponds to the armchair direction, whereas the (820) reflection corresponds to the zigzag direction. The change in relative intensity of these peaks as the sample is rotated azimuthally from 0˚ to 90˚ suggests an uneven volume fraction of the two IP orientations. To quantify these fractions, the intensity associated with each reflection was determined by integrating the counts within an elliptical region surrounding the peak. The intensities were normalized by $|F|^2$, which is proportional to intensity, to account for differences in structure factor between the (820) and (802) reflections. Using this method, we estimate a 4:1 volume fraction between armchair and zigzag orientations for both SnSe and GeSe at φ=0˚. When the sample is rotated 90˚ azimuthally, the grains that contributed to the (802) reflection at φ=0˚ will instead contribute to the (820) reflection as illustrated schematically in Figure S1. Consequently, at φ=90˚we find a reversed ratio of 1:4 corroborating our estimate.

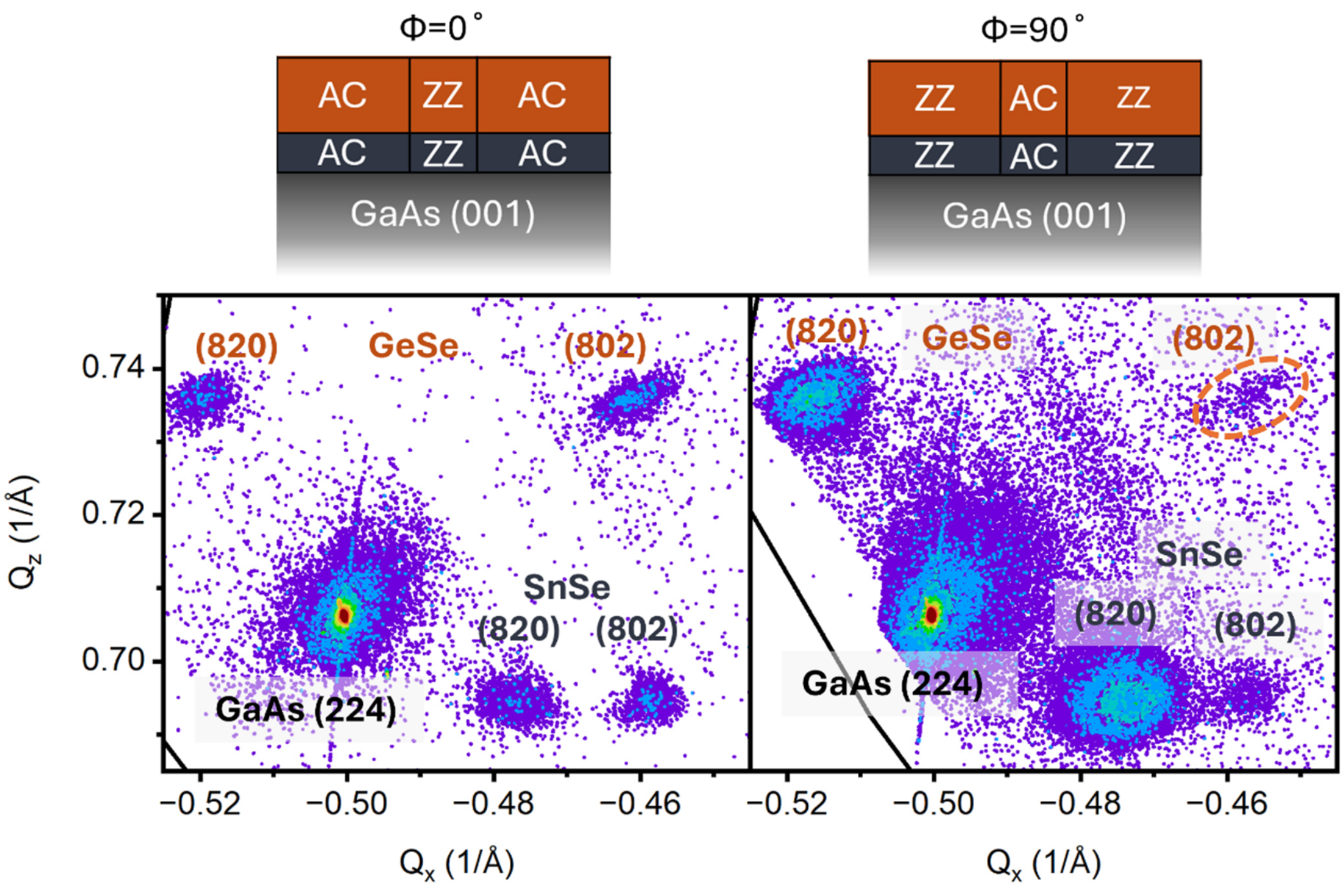

Figure S1. RSMs of GeSe film with SnSe buffer grown on GaAs (001) post annealing at azimuths φ=0˚ and 90˚. Schematics highlight how the GeSe layer maintains the IP orientation of the SnSe buffer after crystallization and the difference in volume fraction (4:1) between the armchair (AC) and zigzag (ZZ) IP orientations.

**S2: Polarized Optical Microscopy of GeSe**

The IP orientation of the SnSe and GeSe layers can be spatially visualized using polarized optical microscopy in reflection mode cross-polarized geometry. In the as-grown sample, the observed optical contrast is from the crystalline SnSe buffer. The two families of blue and yellow grains correspond to the two IP orientations of SnSe. The optical contrast is a result of the birefringence (Δn) and dichroism (Δk) between the armchair and zigzag directions. Once the sample is annealed, the hue of both families of grains changes as the optical response is now dominated by the crystallized GeSe layer.

Notably, the grain structure of the as-grown and annealed samples remains largely unchanged. This suggests that GeSe crystallizes with a high nucleation density during annealing and maintains the local IP orientation of the underlying SnSe buffer. Such a mechanism would produce similar variant populations in the SnSe buffer and GeSe film. This interpretation is consistent with XRD measurements that indicated a 4:1 volume fraction of the two IP orientations in both the SnSe and GeSe layers.

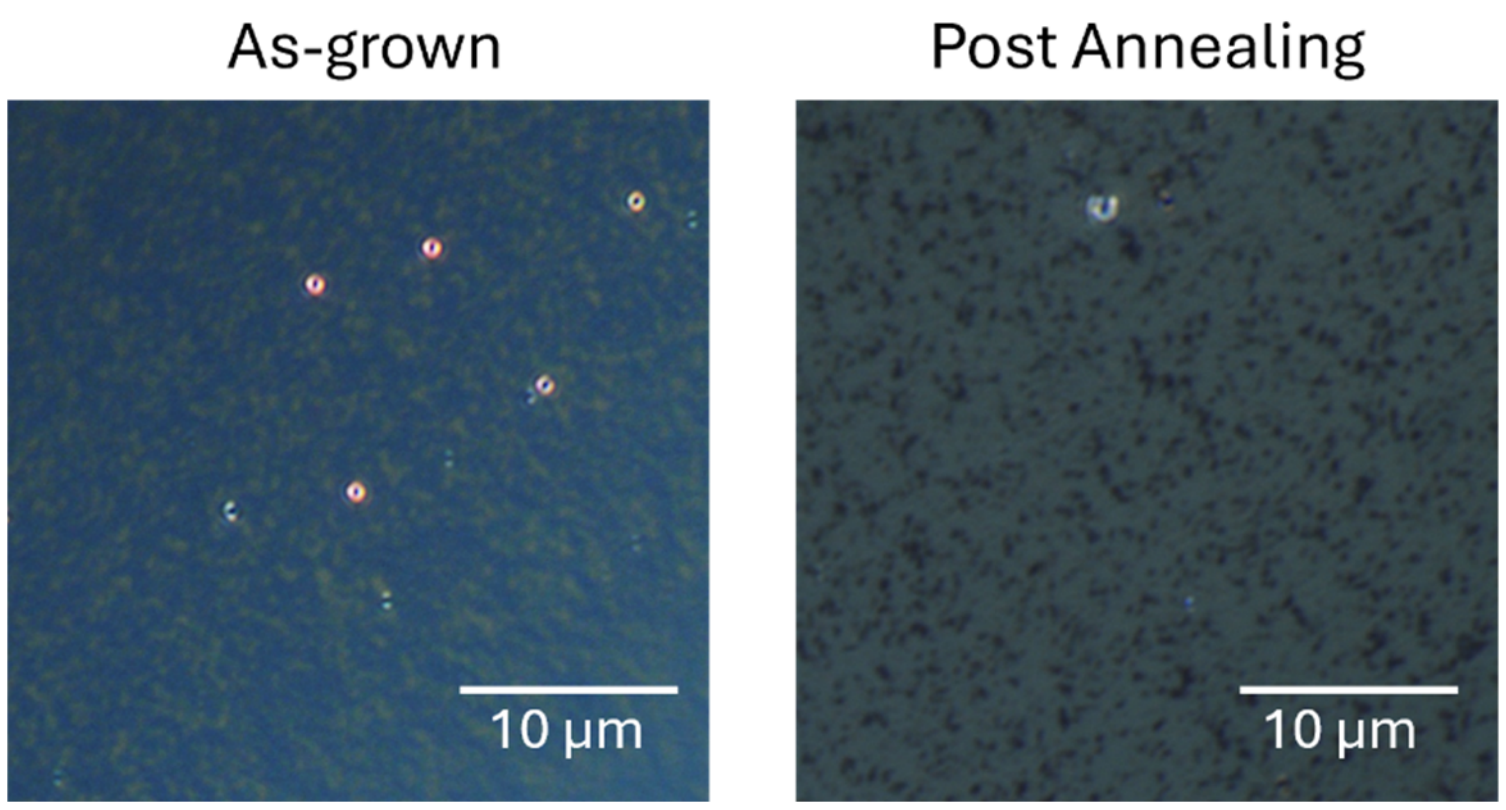

Figure S2. Polarized optical microscope images of as-grown and annealed GeSe with a SnSe buffer grown on on-axis GaAs.

**S3: GeSe 1030 nm polarized reflection**

At 1030 nm, the anisotropy ratio of on-axis and offcut GeSe are measured to be 0.04 and 0.06 respectively. This corresponds to a 1.4× increase in the anisotropic response by stabilizing a single-IP-orientation on an offcut substrate. The expected gain in optical anisotropy resulting from suppressing the minority variant population can be estimated using the armchair and zigzag reflected intensities measured from the single-orientation offcut sample. The first expression corresponds to the fully single-variant film, while the second assumes a 4:1 ratio of two IP orientations, consistent with the volume fractions determined by XRD.

$$anisotropy\ ratio_{1:0} = \frac{\mathrm{I_{ZZ}} - \mathrm{I_{AC}}}{\mathrm{I_{AC}}}$$

$$anisotropy\ ratio_{4:1} = \frac{(0.8\mathrm{I_{ZZ}} + 0.2\mathrm{I_{AC}}) - (0.2\mathrm{I_{ZZ}} + 0.8\mathrm{I_{AC}})}{0.2\mathrm{I_{ZZ}} + 0.8\mathrm{I_{AC}}}$$

From these equations, the expected increase is 1.7×. The experimentally measured increase of 1.4× is slightly smaller, likely due to experimental uncertainties such as small sample misalignment during measurements which can reduce the difference between the maximum and minimum intensities.

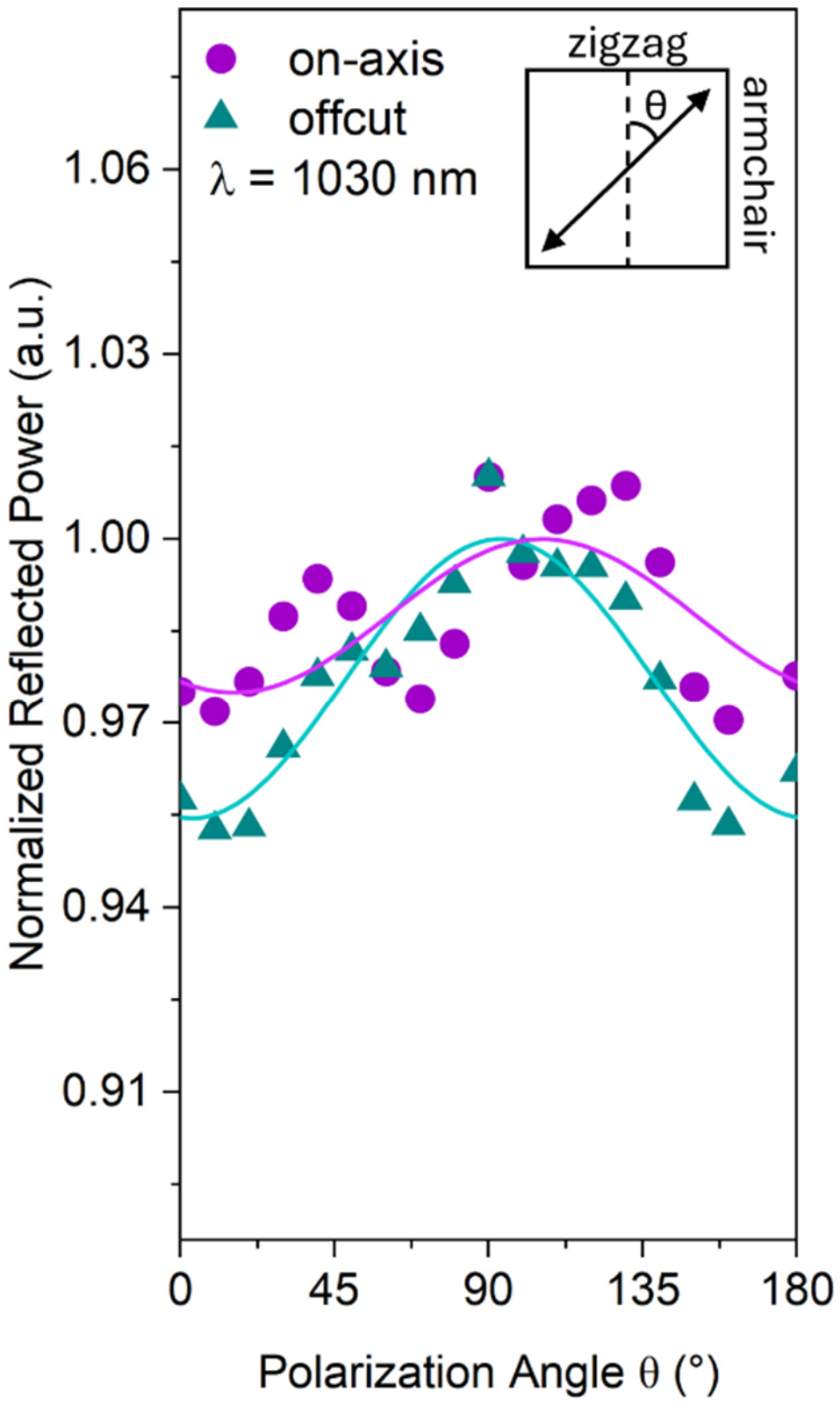

Figure S3. Normalized reflected power vs polarization angle for on-axis and offcut GeSe at 1030 nm.

**S4: Off-(100) Orientations observed in SnSe and SnGeSe**

We observe a tendency for the (311) out of plane orientation to stabilize when SnSe and SnGeSe are grown directly on GaAs at 160˚C. These off-(100) orientations are present in both as-grown crystalline and thermally crystallized samples. For SnSe grown directly on GaAs at 160˚C, open detector 2θ-ω scans show the (311) orientation. In contrast, at a higher growth temperature of 300˚C only (100) orientations are observed as discussed in the main text. Off-(100) orientations require direct covalent bonds with GaAs and may be forming due to limited kinetics at lower growth temperatures or poor selenium passivation of the substrate.

A similar trend is observed for SnGeSe. As grown crystalline SnGeSe ($x_{Ge} \approx 0.18$) deposited directly on GaAs exhibits the (311) orientation out of plane. We also note the appearance of the (311) peak in solid phase crystallized SnGeSe ($x_{Ge} \approx 0.40$) grown on exact and 4˚ offcut GaAs. Comparison of the 2θ-ω scans indicates that the offcut substrate helps stabilize the desired (100) orientations as we index the (400) peak.

These less desirable orientations can be suppressed using a significantly thick SnSe buffer layer grown at 300˚C. With a 10 nm SnSe buffer, a low intensity (311) peak is still observed, suggesting there are uncoalesced parts of the SnSe buffer allowing the SnGeSe to form direct covalent bonds with GaAs substrate. Increasing the SnSe buffer thickness to 50 nm fully suppresses the (311) orientation as shown in the main text.

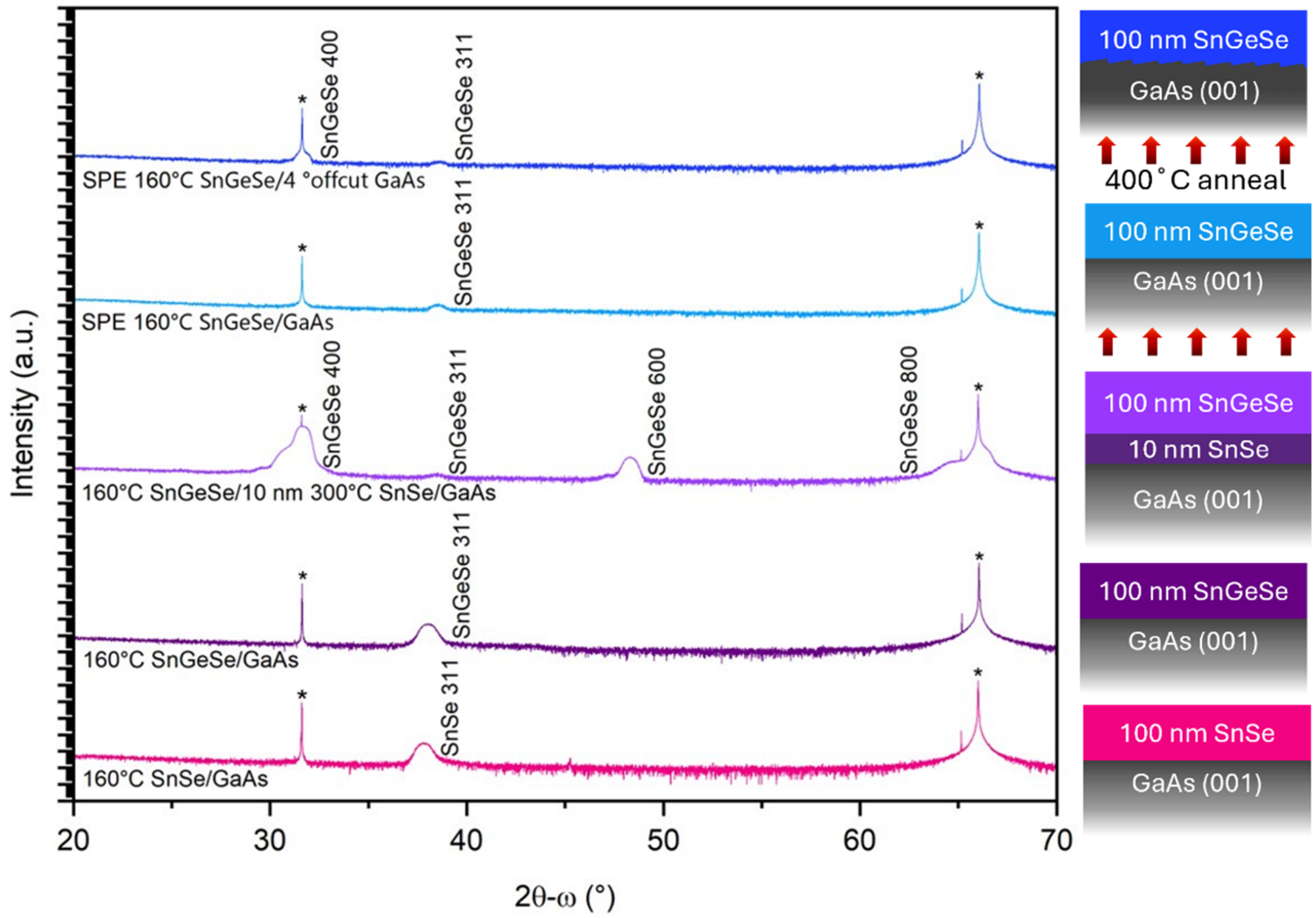

Figure S4. Open detector symmetric 2θ-ω scan and schematics of samples where the (311) orientation was observed. GaAs substrate peaks are denoted with an asterisk.

## S5: Additional Characterization of SnGeSe alloys

Symmetric 2θ-ω XRD scans across the entire alloy space from SnSe to GeSe are shown in Figure S5a. Alloy compositions were estimated using the (800) peak position and bulk alloy lattice constant trends from Krebs.[1] As the Ge content increases, the (800) peak shifts to higher angles as the out-of-plane lattice constant (*a*) decreases. In the pure SnSe film we find the *a* lattice constant is compressively strained by 0.3%. When calculating alloy composition, we assume a constant strain of 0.3% in all films. We also note the $x_{Ge}$=0.45 sample, which was crystallized *ex-situ* has an additional (311) peak. The low intensity of the (311) peak compared to the (400) peak suggests only a small volume fraction of (311)-oriented grains as the structure factors of both reflections are similar in magnitude.

Raman spectroscopy corroborates the change in Ge composition as shown in Figure S5b. Raman spectra were collected at room temperature using a 633 nm laser on the as-grown crystalline alloys. Since SnSe and GeSe share the same *Pnma* structure, they theoretically exhibit the same Raman modes including the $A_{g(1)}$, $B_{3g}$, $A_{g(2)}$, and $A_{g(3)}$.[2] In GeSe, the $A_{g(2)}$ modes are too small to measure experimentally but are detectable in SnSe.[3,4] As the Ge content increases, the Raman peaks seem to generally shift to higher wavenumbers. Additionally, the $A_{g(2)}$ peaks become broader and lower in intensity, eventually becoming indistinguishable from the background in the $x_{Ge}$=0.39 sample.

Furthermore, we plot the Ge content against the ratio of GeSe BEP to total BEP during growth in Figure S5c. The linear trend, despite the increasing growth rate, suggests near unity sticking at the growth temperature of 160˚C. This is ideal for alloy growth as it makes the composition easier to control. When in a non-unity sticking regime, increasing the growth rate while maintaining the same BEP ratio will cause the composition to change. At a BEP ratio of 0.5, increasing the growth rate by a factor of 2× only minimally changes the composition ($x_{Ge}$=0.36 to $x_{Ge}$=0.39 respectively) further supporting near unity sticking.

Open detector rocking curves were used to study how structural quality changes with composition and growth rate in Figure S5d. The FWHM of the pure SnSe sample is around 1780 arcseconds. The FWHM increases as the Ge content and growth rate both increase. To probe the effect of alloying we compare the rocking curves between the $x_{Ge}$=0.0 and $x_{Ge}$=0.36 samples which were grown with the same growth rate. The rocking curve FWHM increases by only 5% from a 36% increase in Ge content. On the other hand, if we compare the $x_{Ge}$=0.36 sample to the $x_{Ge}$=0.39 sample grown at 2× the growth rate, we see the FWHM increases by 20%. This comparison indicates that increasing the growth rate has a stronger effect on the increase in FWHM than the effects of alloying.

We then analyze the rocking curves of our SPE samples separately to study the effect of alloying on solid phase crystallization. The pure GeSe sample has a FWHM of 1700 arcseconds, which is on par with the as-grown crystalline SnSe sample. However, the $x_{Ge}$=0.45 sample has a FWHM of 2940 arcseconds. The large increase in FWHM suggests an increased barrier to epitaxial crystallization in SnGeSe compared to the pure GeSe. This behavior may reflect an increase in the onset crystallization temperature, glass transition temperature, or peak crystallization temperature with alloying. A separate study is needed to better understand the effect of alloying on the solid phase crystallization of these glasses.

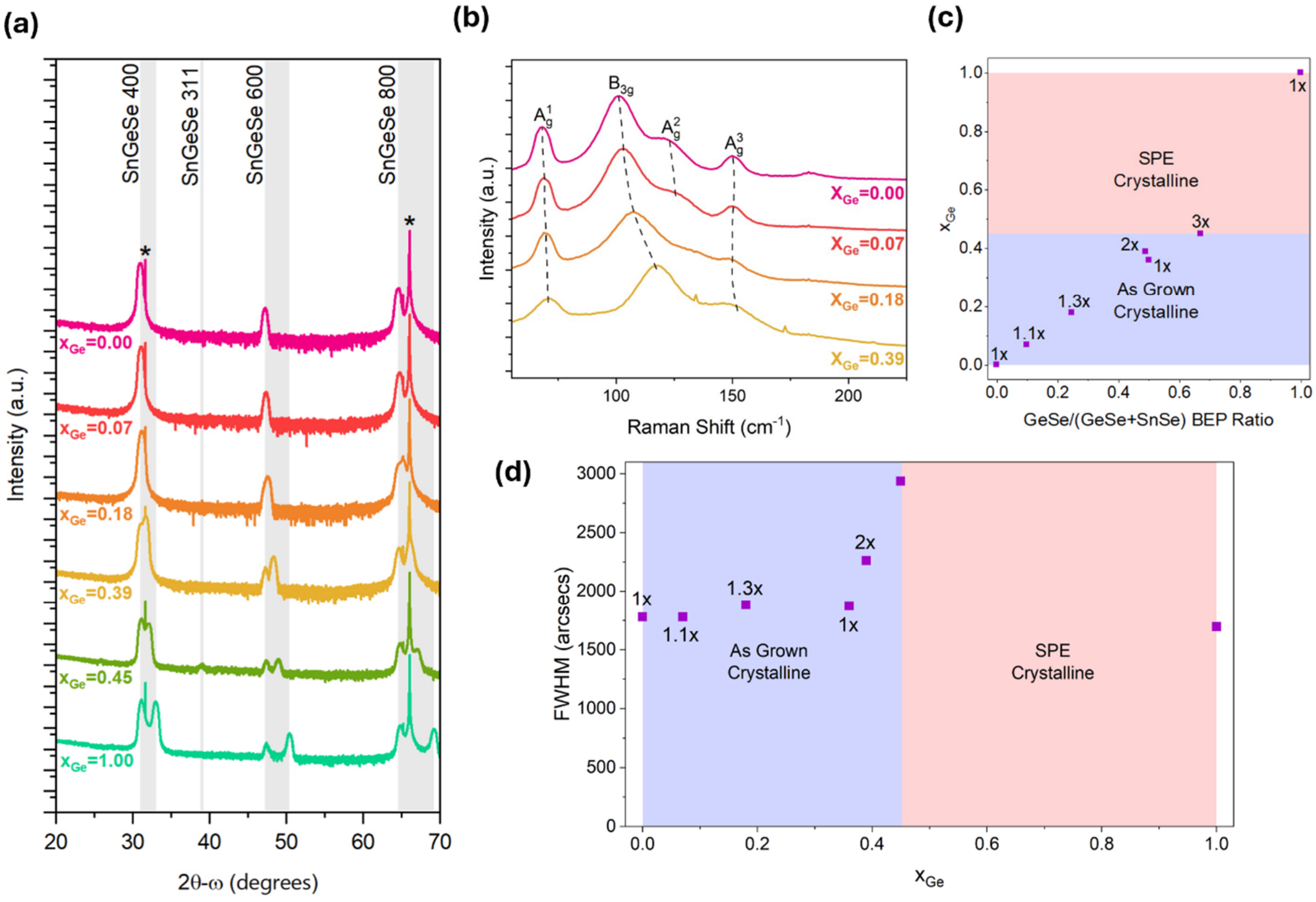

Figure S5. (a) Open detector symmetric 2θ-ω scans across entire alloy space. Alloy compositions were determined from the SnGeSe (800) peak position in triple axis 2θ-ω scans. The asterisk is used to denote GaAs substrate peaks. (b) Raman spectrum of as-grown crystalline SnGeSe alloys. (c) Ge composition vs ratio of GeSe BEP to total BEP. Multipliers represent scaling of pure SnSe growth rate. (d) Rocking curve FWHM of as grown crystalline alloys and SPE crystalline alloys.

**S6: Stabilization of two in-plane orientations on offcut GaAs at low growth temperatures**

Figure S6a shows reciprocal space maps of crystalline SnGeSe ($x_{Ge}$=0.18) grown at 160˚C on top of a 50 nm SnSe buffer deposited at 300˚C on 4˚ offcut GaAs (001). The presence of the SnGeSe (820) peak in both orthogonal azimuths confirms two IP orientations exist despite the buffer layer of SnSe having a single-IP-orientation. Characterizing the fraction of each SnGeSe orientation, we get a 9:2 ratio.

To separate the effects of alloying and growth temperature, we then grew 100 nm of SnSe at 160˚C on a 50 nm SnSe buffer deposited at 300˚C on offcut GaAs (001). RSMs reveal two IP orientation of SnSe as we observe both the (820) and (802) peaks at azimuths φ=0˚ and 90˚ as shown in Figure S6b. Since we have previously shown 300˚C SnSe grown directly on offcut GaAs results in a single-IP-orientation, the double-variant population must form only in the 160˚C SnSe layer.

To confirm the observed effect is unrelated to film thickness, we grew 150 nm of SnSe directly on offcut GaAs at 300˚C. As seen in Figure S6c, the film retains a single-IP-orientation. Together, these results suggest that the stabilization of a second IP orientation is due to low growth temperatures rather than alloying or film thickness. We suspect that at lower growth temperatures the energy barrier between forming the zigzag and armchair orientation is low such that there is no longer a strong preferred orientation. Post-growth annealing had no effect on the IP orientation of the alloy.

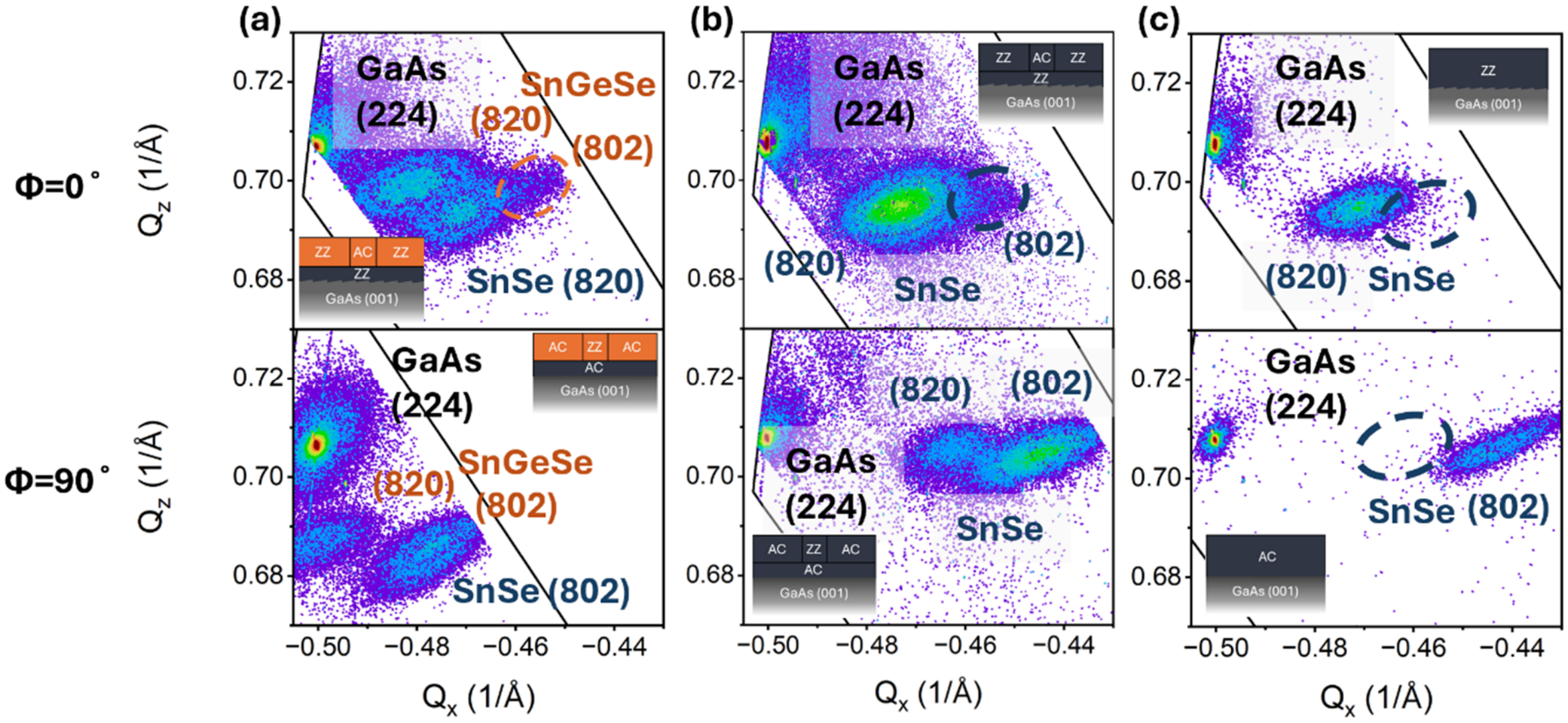

Figure S6. RSMs taken at azimuths φ=0˚ and 90˚ of (a) 160˚C SnGeSe ($x_{Ge}$=0.18) with a 300˚C SnSe buffer, (b) 160˚C SnSe with a 300˚C SnSe buffer, and (c) 300˚C SnSe grown on 4˚ offcut GaAs (001). Dashed ovals mark the expected peak position of the orthogonal in-plane orientation.